\documentclass[aps,prl,twocolumn,groupedaddress,nofootinbib]{revtex4}
\usepackage{graphicx,amsfonts,amsmath,color,epsfig,epstopdf}
\newcommand{\SU}[1]{\ensuremath{\mathrm{SU}( #1 )}}

\newcommand{\SpR}[1]{\ensuremath{\mathrm{Sp}( #1,\mathbb{R} )}}
\newcommand{\etal}{\emph{et al.}}

\newcommand{\half}{\ensuremath{\textstyle{\frac{1}{2}}}}

\newcommand{\betb}{\begin{tabular}{p{4.0cm}p{9.0cm}}}
\newcommand{\entb}{\end{tabular}}

\newcommand{\ho}{\ensuremath{\hbar\Omega}}
\newcommand{\ph}[1]{\ensuremath{#1}p-\ensuremath{#1}h}
\begin{document}

\title{
Hoyle state and rotational features in Carbon-12 \\
within a no-core shell model framework
}
\author{Alison C. Dreyfuss}\affiliation{Keene State College, Keene, New Hampshire 03435, USA}
\author{Kristina D. Launey}
\author{Tom\'a\v{s} Dytrych}
\author{Jerry P. Draayer}
\affiliation{Department of Physics and Astronomy, Louisiana State
University, Baton
Rouge, LA 70803, USA}
\author{Chairul Bahri}
\affiliation{Department of Physics, University of Notre Dame,
Notre Dame, Indiana 46556-5670, USA}

\begin{abstract}
By using only a fraction of the model space extended beyond current no-core shell-model limits and a  many-nucleon interaction with a single parameter, we gain additional insight within a symmetry-guided shell-model framework, into the many-body dynamics that gives rise to the ground state rotational band together with phenomena tied to alpha-clustering substructures in the low-lying states in $^{12}$C, and in particular, the challenging Hoyle state and its first $2^+$ and $4^+$ excitations. For these states, we offer a novel perspective emerging out of no-core shell-model considerations,  including a discussion of associated nuclear deformation and matter radii. This, in turn,  provides guidance for {\it ab initio} shell models by informing key features of nuclear structure and the interaction.
\end{abstract}

\maketitle

Our present-day knowledge of various phenomena of astrophysical significance, such as nucleosynthesis, the evolution of primordial stars in the Universe, and X-ray bursts  depends on reaction rates for the stellar triple-$\alpha$ process, which can considerably affect, e.g., results of core-collapse supernovae simulations and stellar evolution models, predictions regarding X-ray bursts,  as well as estimates of carbon production in asymptotic giant branch (AGB) stars \cite{Fynbo05}. These rates, in turn, are greatly influenced by accurate measurements and theoretical predictions of several important low-lying states in $^{12}$C, including the second $0_2^+$ (Hoyle) state and its $2^+$ excitation that continues to foster debate in experimental studies \cite{Freer0709,Hyldegaard10,Itoh11,ZimmermanDFGS11,Raduta11,Zimmerman13}. Further challenges relate to the long-recognized  alpha-cluster substructure of these states that has been  explored within cluster-tailored  \cite{ChernykhFNNR07,KhoaCK11} or self-consistent \cite{UmarMIO10} microscopic framework, but has hitherto precluded an accurate -- from first principles ({\it ab initio}) --  no core shell-model  (NCSM) description  \cite{RothLCBN11}. Only recently, first {\it ab initio} state-of-the-art calculations have been attempted 
using lattice  effective field theory (EFT) \cite{EpelbaumKLM11}.

In this letter, we report on a first study of these  phenomena in $^{12}$C within a no-core shell-model framework with essentially no limitation on the number of harmonic oscillator (HO) shells included in the model space. While such model spaces remain inaccessible by {\it ab initio} shell models, we are able to address a long-standing challenge \cite{EllisEngeland7072}, namely, understanding highly-deformed spatial configurations from a shell-model perspective. This is achieved by down-selecting, first, to the most physically relevant nuclear configurations and, second, to pieces of the nucleon-nucleon ($NN$) interaction that enter in commonly used nuclear potentials \cite{Elliott58ElliottH62,BohrMottelson69}.

We emphasize that the goal of the present study is to inform key features of nuclear structure and the interaction, and hence, to provide guidance needed for {\it ab initio} shell model  approaches. To do so, we retain simplicity by focusing on two essential pieces, namely, the long-range part of the central nuclear force and a spin-orbit term,  while excluding from the present analysis various interaction terms,  such as short-range and tensor forces. The latter are indispensable for accurate descriptions, but appear to be of secondary importance to the present study, as suggested by  the reasonably close agreement of  the model outcome with experiment and {\it ab initio} results in smaller spaces.
The outcome further points to the need for simple many-body interactions beyond two-body realistic ones for a description of large-deformation and cluster phenomena, and manifests an indication that achieving {\it ab initio}  descriptions is within the reach of the NCSM. The latter, in turn, will bring forward an  accurate reproduction and reliable prediction of energy spectra and associated transition rates that majorly impact astrophysical studies.

In particular, this study allows one to gain further insight  into the many-body dynamics, including the physically relevant  deformation and particle-hole configurations, that gives rise to the ground state ($g.st.$) rotational band (the lowest $0^+$, $2^+$ and $4^+$ states) together with  low-lying states suggested to have a cluster structure ($0_2^+$ Hoyle state  and its $2^+$ and $4^+$ excitations),  as well as a third low-lying $0_3^+$ state in $^{12}$C. We focus on excitation energies and other observables such as matter rms radii, electric quadrupole moments and $E2$ transition rates, as well as compare to wavefunctions obtained by {\it ab initio} shell-model calculations using a realistic $NN$ interaction.  
With no parameter adjustment, the present model we find is also extensible to other light nuclei, as demonstrated \cite{LauneyDDTFLDMVB12}, for example, for the $g.st.$ rotational band of $^{8}$Be (and its low-lying $0^+$ states) as well as of $^{22}$Ne and $^{22,24}$Mg.

\noindent
{\bf Symmetry-adapted shell-model framework. --}
We employ the no-core symplectic model (NCSpM) for symmetry-preserving interactions  with \SpR{3} the underpinning symmetry \cite{RosensteelR77}. This symmetry is  found  inherent to nuclear dynamics -- a result we have demonstrated in an analysis of large-scale {\it ab initio} NCSM applications for $^{12}$C and $^{16}$O \cite{DytrychSBDV06}. The model offers a microscopic description of $A$ nucleons in terms of mixed  deformation configurations and associated rotations \cite{RosensteelR80}, directly related to particle relative (with respect to the center of mass, CM) position  and momentum coordinates, ${\mathbf r}_i$ and ${\mathbf p}_i$, with $i=1,\dots,\, A$. It has been successfully applied to $^{20}$Ne \cite{DraayerWR84} with a $^{16}$O core, as wells as to  $^{166}$Er using the Davidson potential \cite{BahriR00}.  It is a microscopic realization of the Bohr-Mottelson collective model \cite{BohrMottelson69}, as well as a multiple HO shell generalization of Elliott's \SU{3} model \cite{Elliott58ElliottH62}.

The NCSpM utilizes a symplectic basis (for details, see \cite{DytrychSDBV08}), which is related -- via a unitary transformation -- to the three-dimensional HO ($m$-scheme) many-body basis used in the NCSM \cite{NCSM}. The NCSM basis is constructed using HO single-particle states. It is characterized by the $\hbar\Omega$ oscillator strength 
and  by the cutoff in total excitation oscillator quanta, $N_{\max} $. Indeed, the NCSpM employed within a full model space up through $N_{\max}$, will coincide with the NCSM for the same $N_{\max}$ cutoff. It is therefore clear that the present study, while down-selecting to the most relevant configurations, provides the first shell-model calculations carried beyond current NCSM limits. These important configurations are chosen among all possible symplectic \SpR{3} irreducible representations (irreps) within the model space. 
\begin{figure}[th]
\begin{center}
\includegraphics[width=0.45\textwidth]{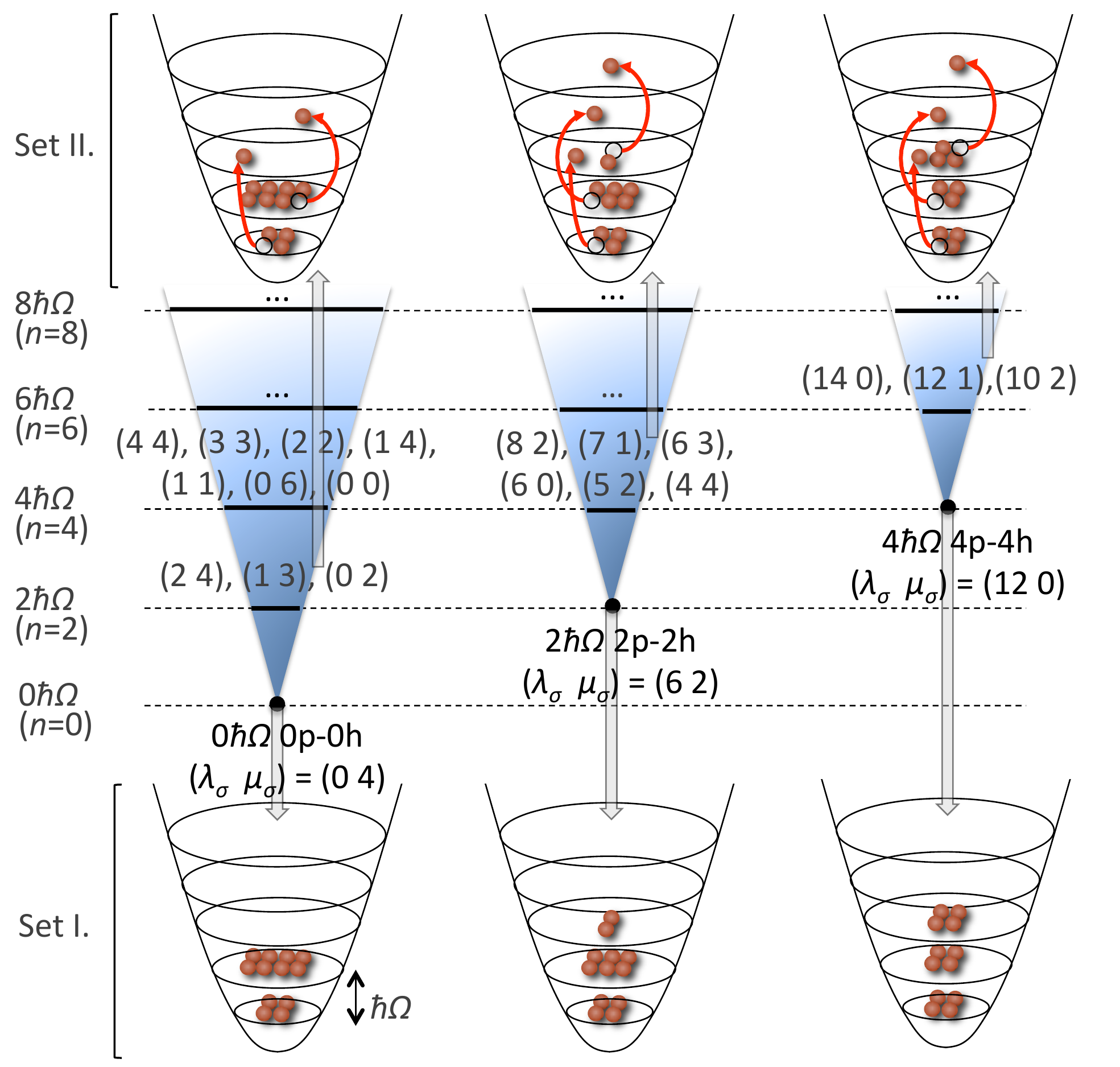} 
\end{center}
\caption{
\SpR{3} irreps (slices) that comprise the spin-zero model space used for the $^{12}$C NCSpM calculations. Basis states $(\lambda\,\mu)$ of a slice are built by 2\ho~\ph{1} monopole or quadrupole excitation (Set II) over a bandhead. The symplectic  bandhead (Set I) is a \SU{3}-coupled many-body state with a given nucleon distribution over the HO shells. The corresponding HO energy of this nucleon configuration
together with the bandhead deformation, $(\lambda_\sigma\,\mu_\sigma)$, serve to label the symplectic irrep.}
\label{Sp3r_picture}
\end{figure}

The \SpR{3} irreps divide the space into `vertical slices' that are comprised of basis states of definite $(\lambda\,\mu)$ quantum numbers of \SU{3} (Fig. \ref{Sp3r_picture}) linked to the intrinsic quadrupole deformation \cite{RosensteelR77bLeschberD87CastanosDL88}. E.g., the simplest cases, $(0\, 0)$,
$(\lambda\, 0)$, and $(0\,\mu)$,  describe spherical, prolate, and oblate deformation, respectively, while a general nuclear state is typically a superposition of several hundred various triaxial deformation configurations.  The basis states are built over a bandhead (Fig. \ref{Sp3r_picture}, Set I) by  consecutive 2\ho~ \ph{1} ($1$-particle-$1$-hole) excitations (Fig. \ref{Sp3r_picture}, Set II), together with a smaller 2\ho~\ph{2} (two particles a shell up) correction for eliminating the spurious CM motion (not shown in the figure). In the NCSpM, to eliminate the spurious CM motion, we use 
symplectic generators constructed in relative coordinates with respect to the  CM. These generators are used to build the basis, the interaction, the many-particle kinetic energy operator, as well as to evaluate observables.

For the purposes of this study, we utilize a microscopic many-body interaction suitable for large-$N_{\max}$ no-core shell model applications.  Specifically, along with the usual spin-orbit term, we employ an elementary form tied to a long-range expansion of the nucleon-nucleon central force $V(|{\mathbf r}_i-{\mathbf r}_j|)$ \cite{Harvey68} kept  
as simple as possible by considering the most relevant degrees of freedom for a description of deformed spatial configurations \cite{Elliott58ElliottH62,BohrMottelson69}, 
\begin{equation}
H_\gamma = \sum_{i=1}^A \left(\frac{{\mathbf p}_{i}^2}{2m}+\frac{m\Omega^2 {\mathbf r}_{i}^2}{2} \right) + \frac{\chi}{2} \frac{\left( e^{-\gamma Q\cdot Q} -1 \right)}{\gamma} -\kappa \sum_{i=1}^A l_i\cdot s_i.
\label{effH}
\end{equation}
This Hamiltonian is given in terms of particle coordinates relative to the CM, with $Q_{(2\mathcal{M})}
=\sum_{i=1}^Aq_{(2\mathcal{M})i}
=\sum_i \sqrt{16\pi/5} r_i^2Y_{(2\mathcal{M})}(\hat {\mathbf r}_i)$ the mass quadrupole moment and with  $\half Q\cdot Q = \half \sum_{i} q_i\cdot (\sum_j q_j)$ the interaction of each particle with the total quadrupole moment of the system\footnote{
Although a technical detail, it is important to note that  $Q\cdot Q$ in (\ref{effH}) denotes the $Q\cdot Q-\langle Q\cdot Q\rangle_n$ interaction, where the $\langle Q\cdot Q\rangle_n$ term is subtracted from $Q\cdot Q$ to eliminate a spurious shift in the zero-point energy by the average contribution, $\langle Q\cdot Q\rangle_n$, of $Q\cdot Q$ within the subspace of $n$ HO  excitations \cite{RosensteelD85,CastanosD89}.
}.
Interactions  that preserve the \SpR{3} symmetry, such as $H_\gamma$ with no spin-orbit term, do not mix symplectic vertical slices, as well as greatly facilitate the use of a group-theoretical apparatus and analytical expressions for the Hamiltonian matrix elements, which, in turn, renders large $N_{\max}$ spaces manageable.
Indeed, \SpR{3}-symmetric Hamiltonians are particularly suitable to capture the essential characteristics of the low-energy nuclear kinematics and dynamics.
The reason is that such Hamiltonians can include 
the many-particle kinetic energy, $\sum_{i} {{\mathbf p}_{i}^2/(2m)}$, the 
HO potential, $\sum_{i} m\Omega^2 {\mathbf r}_{i}^2/2$, as well as terms dependent on $Q$ and on the orbital momentum, ${\mathbf L}=\sum_i {\mathbf r}_i \times {\mathbf p}_i$.

The $H_\gamma$ introduces simple but important many-body interactions that enter in a prescribed hierarchical way given in powers of a small positive parameter $\gamma$, the only adjustable parameter in the model. 
The NCSpM, with $H_\gamma$,  reduces to the established Elliott model in the limit of a single valence shell and zero $\gamma$ and $\kappa$, where it was shown to effectively describe rotational features of light nuclei \cite{Elliott58ElliottH62}. 
A successful extension to  multiple shells has been achieved and applied to the $^{24}$Mg $g.st.$ rotational band \cite{PetersonH80}, where 
an interaction given as a polynomial in $Q$ up through $(Q\cdot Q)^2$ was employed. Such an interaction directly ties to our effective Hamiltonian (\ref{effH}). Indeed, while higher-order terms in $Q\cdot Q$ of Eq. (\ref{effH}) account for  a renormalization of the $\chi$ coupling constant as shown in Ref.  \cite{LeBlancCVR86},
they become quickly negligible  for a reasonably small $\gamma$ and, e.g., for $^{12}$C, we find only one term (three terms), besides $Q\cdot Q$, to be sufficient for the $g.st.$ (Hoyle state). 
\begin{figure}[t]
\begin{center}
\includegraphics[height=0.23\textwidth]{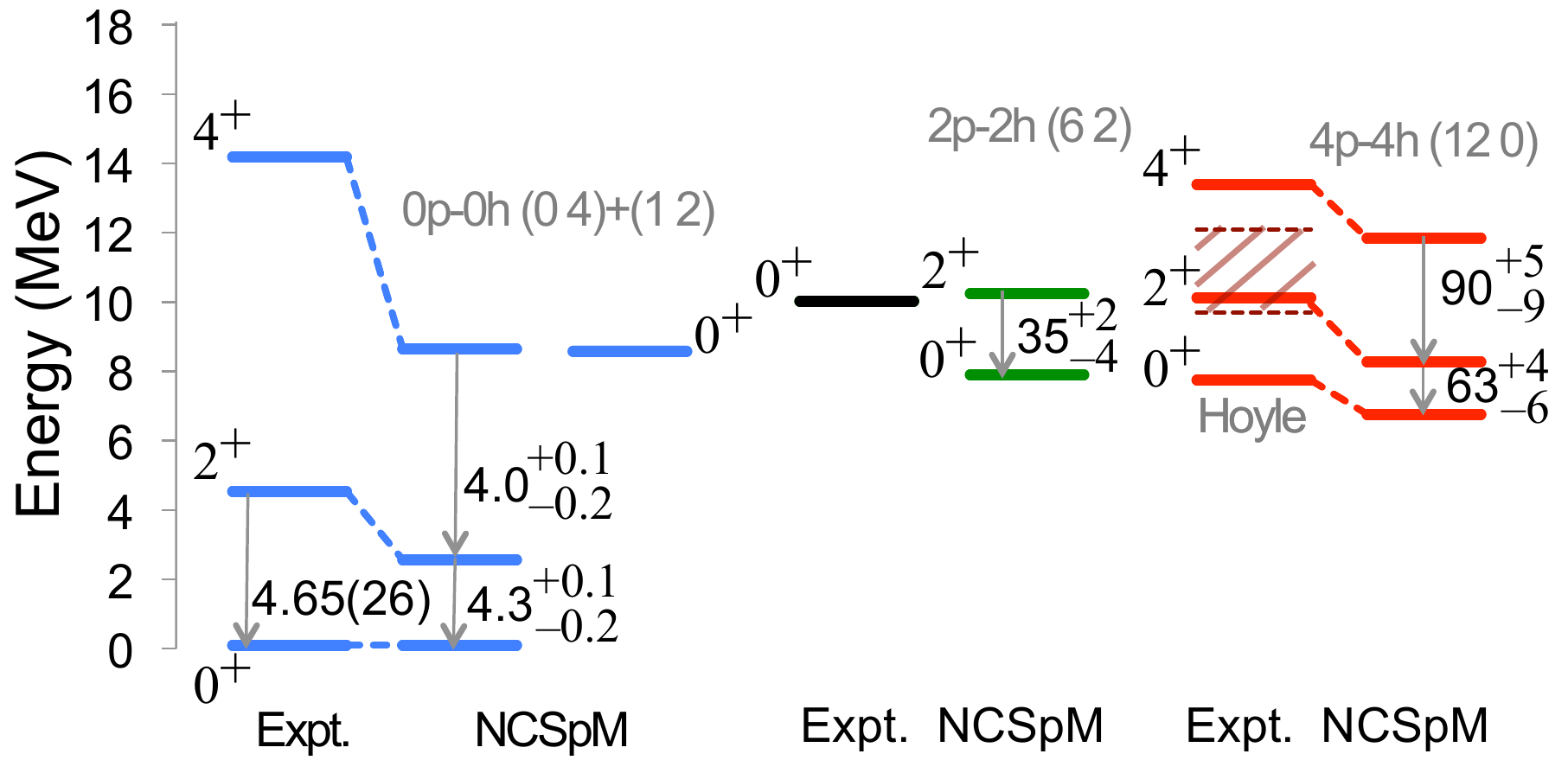} 
\end{center}
\caption{
NCSpM energy spectrum of $^{12}$C. 
Experimental data  is from \cite{ASelove90}, except the latest results for  $0_3^+$ \cite{Itoh11} and the states above the Hoyle state, $4^+$  \cite{Freer11} and $2^+$  \cite{Zimmerman13} (with a shaded area showing the  energy range from \cite{Freer0709,Hyldegaard10,Itoh11,ZimmermanDFGS11,Raduta11}). 
$B(E2)$  transition rates are in W.u.; theoretical uncertainties are estimated for a $\pm 60\%$ deviation of the Hoyle state energy.
}
\label{enSpectrumC12}
\end{figure}

\noindent
{\bf Collective features in $^{12}$C. -- } 
The low-lying energy spectrum and eigenstates for $^{12}$C were calculated using the NCSpM with $H_\gamma$ of Eq. (\ref{effH}) with the empirical estimates $\ho \approx 41/A^{1/3}=18$ MeV and $\kappa \approx 20/A^{2/3}=3.8$ MeV (e.g., see \cite{BohrMottelson69}). We fix the coupling constant $\chi$ by \ho~ and take it to decrease with increasing $N$, the total HO quanta,
by using the relation derived by Rowe \cite{Rowe67}, $\chi=\frac{\ho}{4 N}$ to a leading order. This derivation for $\chi$ is  based on self-consistent arguments and is used in an \SpR{3}-based study of  cluster states of $^{16}$O \cite{RoweTW06}.  

The results are shown for $N_{\max}=20$, which we found sufficient to yield convergence. This $N_{\max}$ model space is further reduced by selecting the most relevant symplectic irreps of total dimensionality of $6.6\times 10^3$.  These irreps (vertical slices) include symplectic excitations that start at the  $0\ho$ \ph{0} $(0\, 4)$, $2\ho$ \ph{2} $(6\, 2)$, and  $4\ho$ \ph{4} $(12\, 0)$ bandheads of  total spin $S=0$ (proton and neutron spins $S_{\rm p,n}=0$) (Fig. \ref{Sp3r_picture}), as well as two $0\ho$ \ph{0} $(1\, 2)$ bandheads of $S=1$ with $S_{\rm p(n)}=0 (1) $ and $1 (0) $. 
In comparison to the experimental energy spectrum (Fig. \ref{enSpectrumC12}), the outcome reveals that, for $\gamma =1.7\times 10^{-4}$, the \ph{0} and \ph{4} symplectic irreps track with the  rotational bands of the $g.st.$ and Hoyle state, respectively, while the lowest \ph{2} $0^+$ is found to lie above the other two $0^+$ states, close to the 10-MeV $0^+$ resonance observed in $^{12}$C. 
Clearly, we find no other  $0^+$ state lying below the calculated Hoyle state for the $\gamma$  chosen, even if all 0\ho~symplectic irreps were included in the $N_{\max}=20$ model space. And further, e.g., for $2\ho$ $(2\,4)$, $4\ho$ $(8\,2)$, $(4\,4)$, and $(0\,6)$, $6\ho$ $(14\,0)$, as well as $8\ho$ $(16\,0)$, the lowest  state ($0^+$) lies much higher than 30 MeV. 
We note that the symmetry-mixing spin-orbit term is turned on only for the symplectic bandheads (up through  $N_{\max}=8$ in this study).
Neglecting the spin-orbit force results in a rather compressed spectrum. This is similar to the findings of early cluster models, which remedy this by allowing for alpha-cluster dissociation due to a spin-orbit force as discussed in Ref. \cite{KanadaEnyo98}. 
\begin{table}[t]
\caption{ NCSpM point-particle rms matter
radii and electric quadrupole moments for $^{12}$C compared to experimental (experimentally deduced) data. See text for a comparison to $r_{\rm rms}$ predictions  of other models. }
{\footnotesize
\begin{tabular}{l|llll}
\hline
 & \multicolumn{2}{c}{matter   $r_{\rm rms}$, fm} & \multicolumn{2}{c}{$Q$, $e\,$fm$^2$} \\
&  Expt. & NCSpM 	 &  Expt. & NCSpM \\ 
	\hline
$0^+_{gs}$  & $2.43(2)$\begin{tablenote}
{\scriptsize Ref. \cite{Tanihata85};  $^b$Ref. \cite{DanilovBDGO09}; $^c$Ref. \cite{Ogloblin13}; and $^d$Ref. \cite{ASelove90}.\\
*Experimentally deduced, based on model-dependent analyses of diffraction scattering; $0^+_{gs}$ $r_{\rm rms}= 2.34$ fm.}
\end{tablenote}
 & $2.43(1)$ & & \\
$0^+_{2}$  {\scriptsize (Hoyle)}  & $2.89(4)^{b*}$  & $2.93(5)$ & & \\
$0^+_{3}$ &  N/A & $2.78(4)$  & & \\
$2_1^+$  & $2.36(4)^{b*}$ & $2.42(1)$  & $+6(3)^{d}$ & $+5.9(1)$\\
 $2^+$ above $0^+_{2}$  & $3.07(13)^{c*}$ & $2.93(5)$  &  N/A  & $-21(1)$\\
$4_1^+$  & N/A & $2.41(1)$  & N/A & $+8.0(3)$\\
$4^+$ above $0^+_{2}$  & N/A & $2.93(5)$  &  N/A  & $-26(1)$\\
 \hline 
 \end{tabular}
}
\label{C12observable}
\end{table}%

The model with the selected $\gamma$ successfully reproduces other observables for $^{12}$C that are informative of the state structure, such as  $r_{\rm rms}$ point-particle matter rms radii, $Q$ electric quadrupole moments (Table \ref{C12observable}) and $B(E2)$ transition strengths (Fig. \ref{enSpectrumC12}).
Theoretical uncertainties in this study are estimated for a $\pm 60\%$ deviation of the Hoyle state energy. This large tolerance corresponds to only $-8\%$ to 15\% variation in $\gamma$ and, practically, has no considerable effect on these observables, e.g., it results in a 0.3-1.6\% (2.3-10\%) variation for radii ($E2$ transition strengths).
The NCSpM finds a quite reasonable matter rms radius for the $g.st.$ (Table \ref{C12observable}).
 Interestingly, our calculations yield matter $r_{\rm rms}=2.93$ fm for the Hoyle state, $1.2$ times larger than that of the ground state. 
While this result drastically differs from predictions of cluster models, e.g., 3.38 fm  \cite{ChernykhFNNR07},  3.83 fm \cite{FunakiTHSR03} and  4.31 fm \cite{YamadaS}, 
it is close to a recent value deduced from experiment,  $2.89(4)$ fm \cite{DanilovBDGO09}, as well as tracks with the {\it ab initio} lattice EFT results at a leading order, 2.4(2) fm  \cite{EpelbaumKLM11}. 
\begin{figure}[t]
\begin{center}
\includegraphics[width =0.23\textwidth]{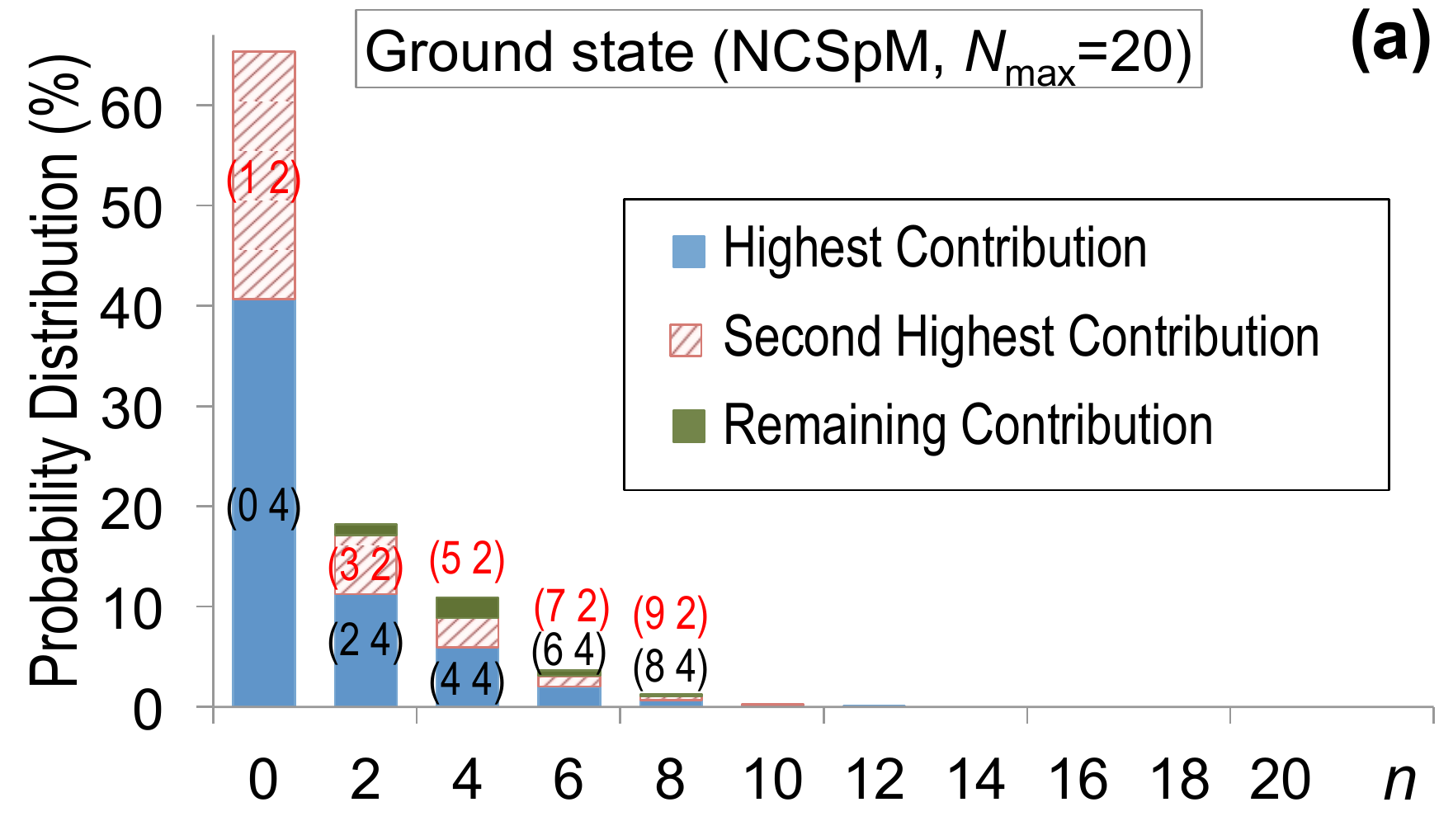}
\includegraphics[width =0.23\textwidth]{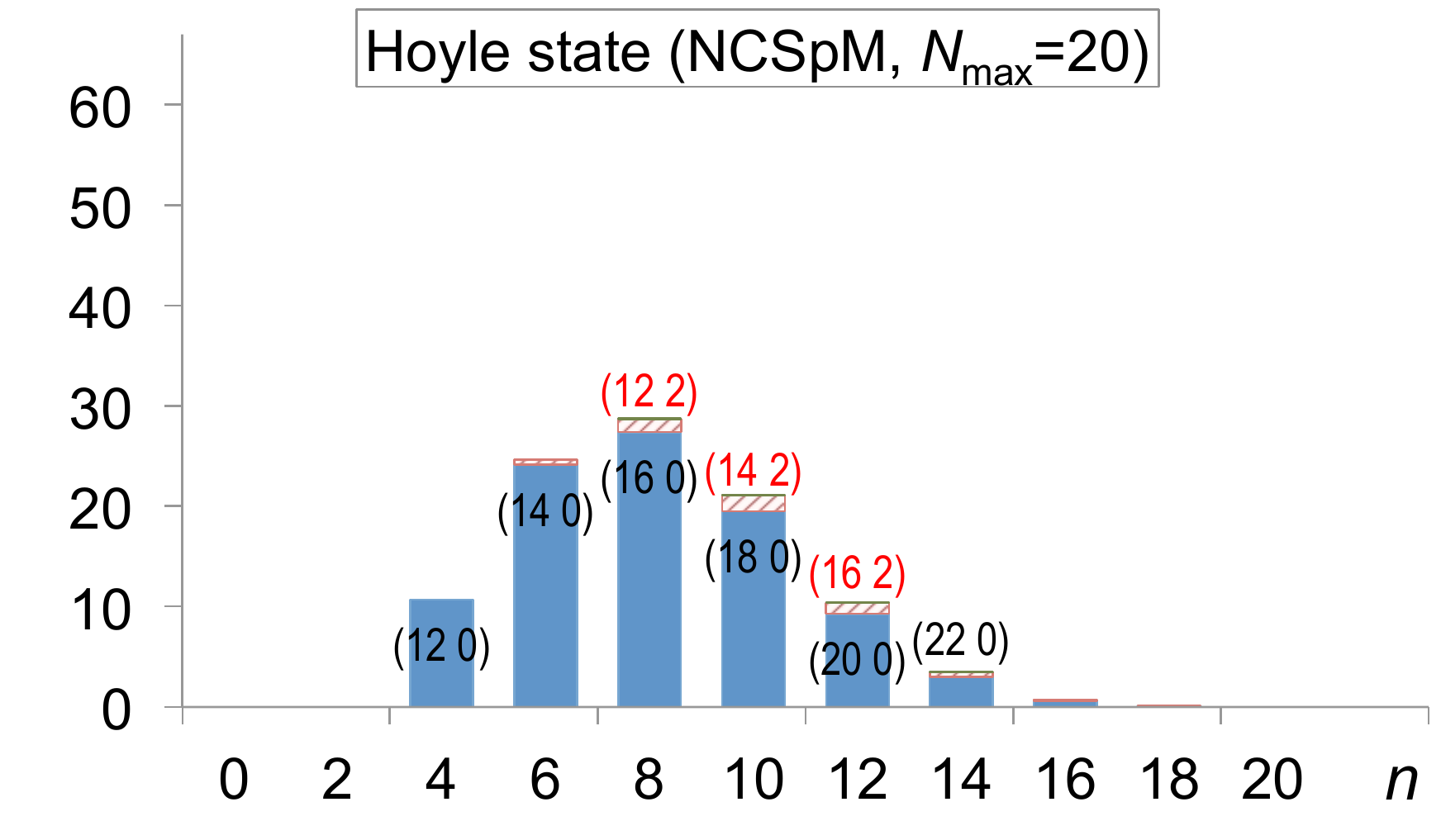} \\
\includegraphics[width=0.23\textwidth]{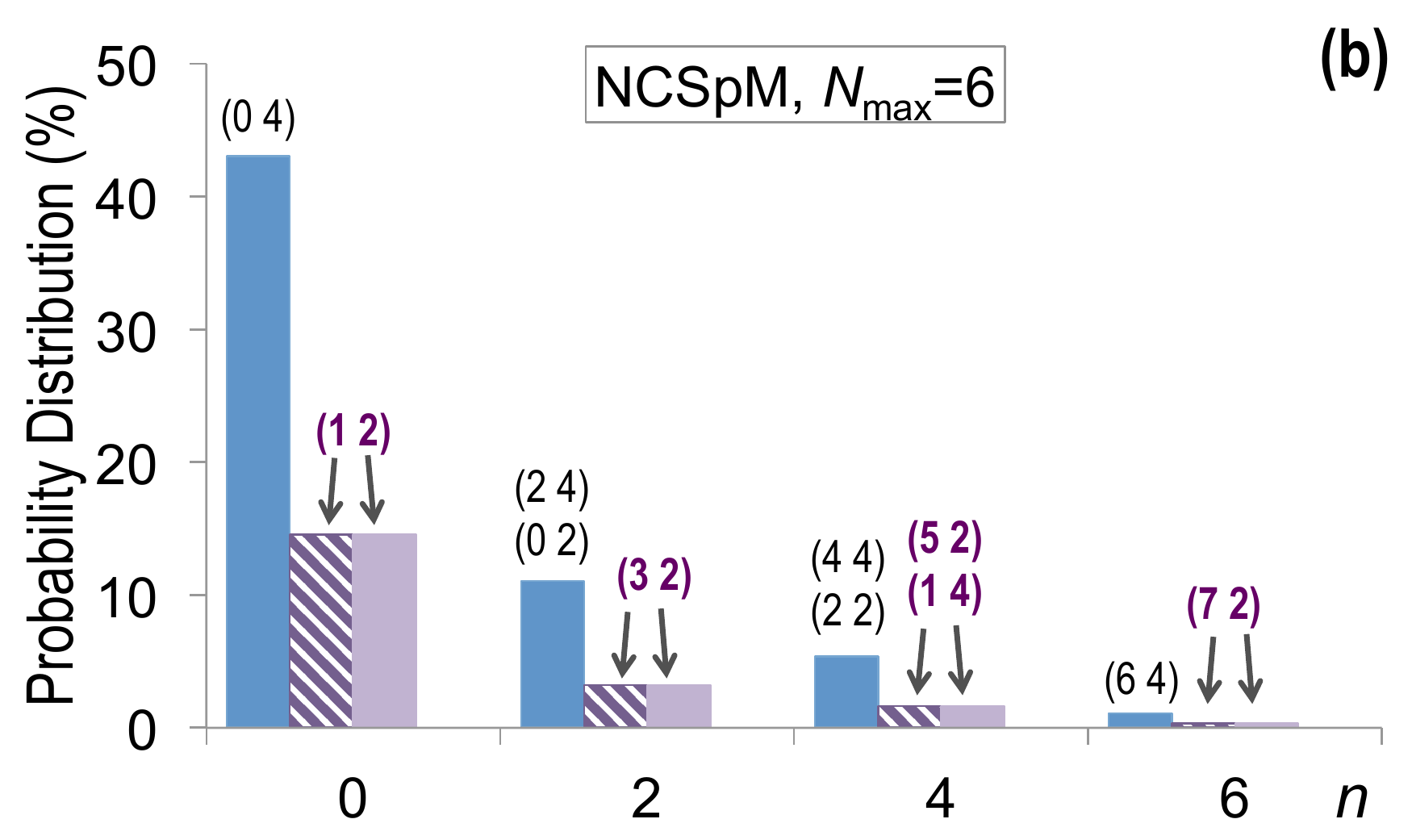}
\includegraphics[width =0.23\textwidth]{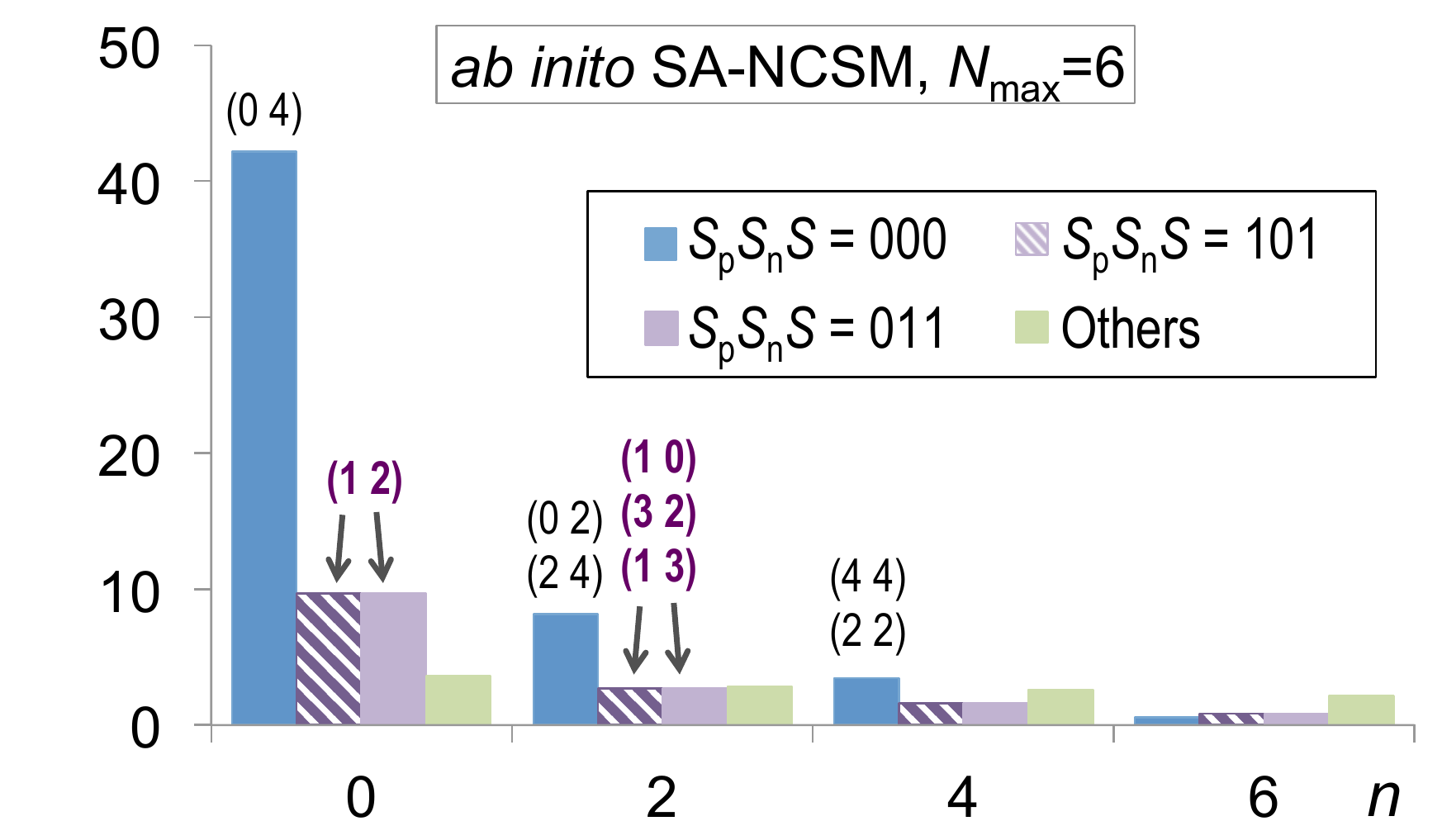}
\caption{\label{wvfnProbabilities}
Probability distribution for $^{12}$C vs. the $n$ total excitations of (a) the ground state (left) and Hoyle state (right) as calculated by NCSpM in $N_{\rm max}=20$, and (b) the
ground state as calculated by NCSpM (left) and SA-NCSM (right) for $N_{\rm max}=6$ and $\ho=18$ MeV. Dominant  deformation modes (with probability amplitude $ \ge 1\%$), specified by $(\lambda\,\mu)$, are also shown.
}
\end{center}
\end{figure}

Furthermore, the model yields a $Q_{2_1^+}$ very close to the experimental value, and a large negative one for the $2^+$ above the Hoyle state (Table \ref{C12observable}) indicating a substantial prolate deformation for the Hoyle and $2^+$ states. Such a deformation, albeit not  so pronounced, has been also suggested  by the {\it ab initio} lattice EFT \cite{EpelbaumKLM11}.
This is also supported by the wavefunction distribution for the $g.st.$ and Hoyle-state rotational bands (see, Fig. \ref{wvfnProbabilities}a for the  lowest $0^+$  states). Namely, while the predominant component of the \ph{0}  states is at 0\ho~($n=0$) and manifests an evident oblate shape [as indicated by the $(\lambda\,\mu)=(0\, 4)$ and $(1\, 2)$ deformation labels], the Hoyle-state band peaks around 8\ho~($n=8$) with a clear evidence for a prolate deformation with $(16\,0)$ being the largest contribution. Moreover, the Hoyle state emerges from a \ph{4} shell-model configuration, that is, the $(12\, 0)$ bandhead (Fig. \ref{Sp3r_picture}, set I) is realized by an alpha-particle configuration -- spatially spherical $(0\,0)$, spin zero, and isospin zero --  in each of the three lowest HO shells (implying spatial displacement). This together with the  strong prolate deformation of this state  supports an underlying alpha-particle cluster structure.

The NCSpM $B(E2;\,2^{+}\!\rightarrow\!0^{+})$ estimates are also found to agree with experiment for the $g.st.$ band,  and with the value of $62.5$ W.u. of Ref. \cite{KhoaCK11} for the Hoyle-state band (Fig. \ref{enSpectrumC12}).
However, nonzero interband $B(E2;\,0^{+}_2\!\rightarrow\!2^{+}_1)$ and $M(E0;\,0^{+}_2\!\rightarrow\!0^{+}_1)$ strengths can only result from mixing of \SpR{3}  irreps, which requires a symmetry-breaking interaction. But this can enter perturbatively, as less than a 2\%  mixing of the $(12\, 0)$ irrep into the $(0\, 4)$ irrep can already yield the correct order of magnitude, namely, $B(E2;\,0^{+}_2\!\rightarrow\!2^{+}_1)=8.4$ W.u. and $M(E0;\,0^{+}_2\!\rightarrow\!0^{+}_1)=2.1$ $e\,$fm$^2$  as compared to experiment, 8.0(11) W.u. and  5.4(2) $e\,$fm$^2$ \cite{ASelove90}, respectively. Estimates of the latter include 6.53 $e\,$fm$^2$ \cite{ChernykhFNNR07}, 3(1) $e\,$fm$^2$ \cite{EpelbaumKLM11}, and 6.7 $e\,$fm$^2$ \cite{KhoaCK11}. 
\begin{figure}[t]
\begin{center}
\includegraphics[width =0.48\textwidth]{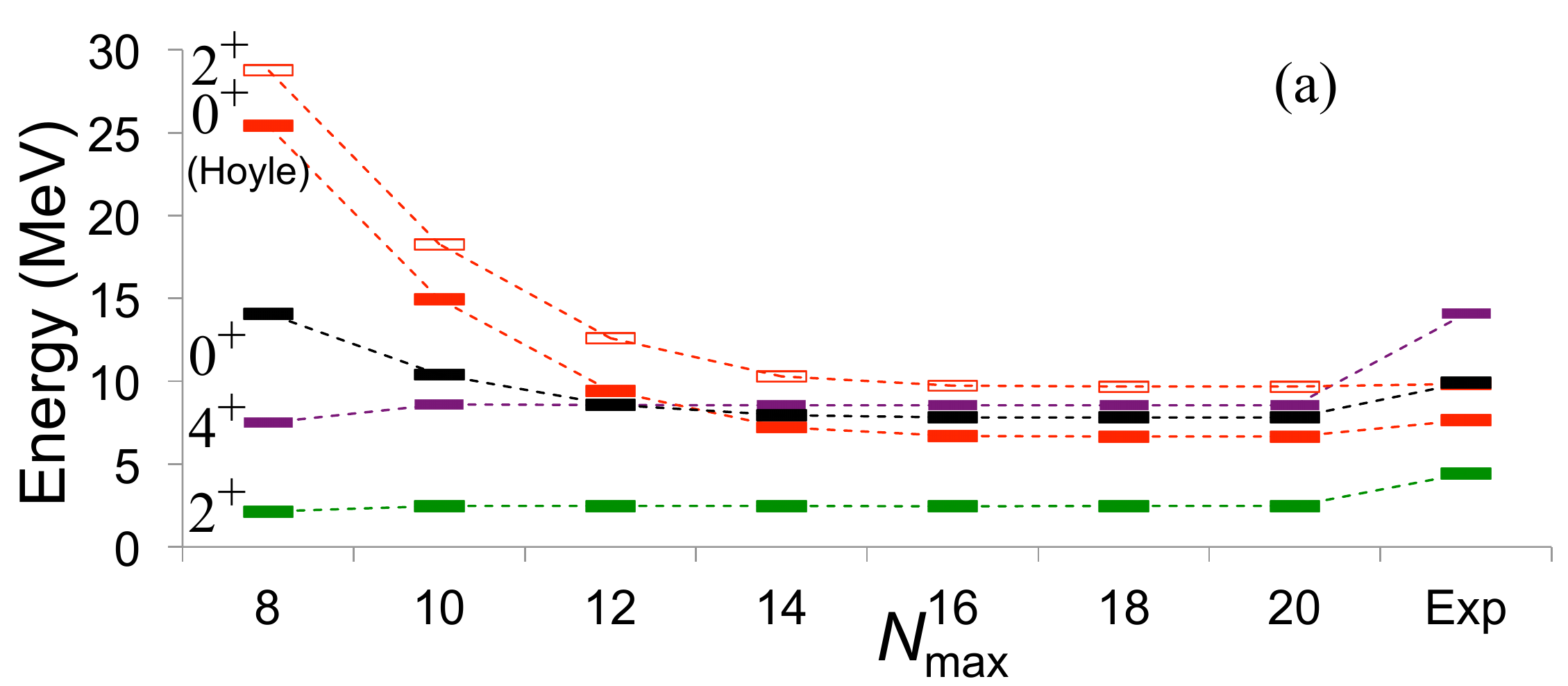}
\includegraphics[width =0.48\textwidth]{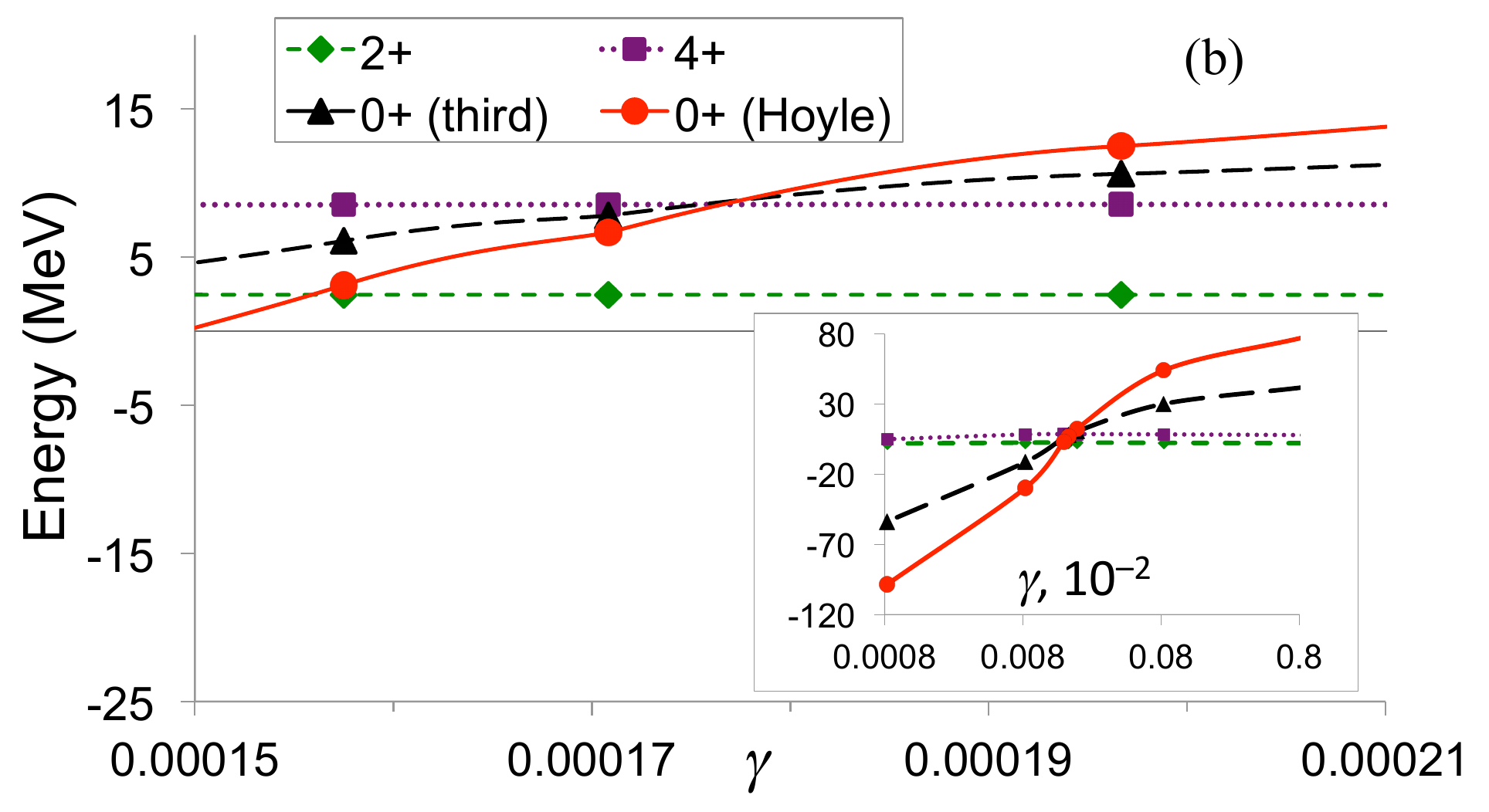}
\end{center}
\caption{
Dependence of the $^{12}$C NCSpM energy spectrum on (a) the model space ($N_{\rm max}$) for  $\gamma=1.7\times 10^{-4}$ and (b) on the $\gamma$ parameter for $N_{\rm max}=20$. }
\label{gammadepend}
\end{figure}

\noindent
{\bf Comparison to {\it ab initio} results. -- } 
A close similarity is observed when  the NCSpM wavefunctions of the $g.st.$ rotational band are compared to {\it ab initio} results for the same $\ho=18$  and $N_{\max}=6$ model space (Fig. \ref{wvfnProbabilities}b). This space appears to be reasonable for these states for both models. In particular, we compare to wavefunctions obtained in the symmetry-adapted no-core shell model (SA-NCSM) \cite{Dytrych13} with bare JISP16 realistic interaction \cite{ShirokovVMW07}. The SA-NCSM utilizes an \SU{3}-coupled basis, which yields conventional NCSM wavefunctions \cite{NCSM}, but realized in terms of the $(\lambda\,\mu)$ deformation labels. The close agreement shows that among all possible configurations present in the SA-NCSM, only the states of the $(0\, 4)$ and then  $(1\, 2)$ symplectic slices appear dominant. And if the SA-NCSM model space is reduced to only the spin components used in this study, $S_{\rm p}S_{\rm n}S=$ $000$, $011$, and $101$, NCSpM observables as $g.st.$ matter rms radius and $Q_{2^+_1}$  reproduce the {\it ab initio} counterparts as much as 80-90\% and 70-90\%, respectively, for the same $\ho$ and $N_{\max}=2,\,4$ and $6$. This suggests that the interaction used in NCSpM has effectively captured a good portion of the underlying physics of the realistic interaction important to the low-energy nuclear dynamics. 

For the Hoyle state and its rotational band, larger spaces are needed, e.g., nonnegligible configurations extend to $N_{\max}=18$ (Fig. \ref{wvfnProbabilities}a for the $0^+$ state), which is within a reach of next-generation {\it ab initio} NCSM models. 
For comparison, recent {\it ab initio} $N_{\max}=8$ NCSM  calculations, while achieving a remarkable reproduction of the $g.st.$ rotational band, yield the second $0^+$ and $2^+$ states around 13 MeV and 15 MeV, respectively \cite{RothLCBN11}, thus believed not to be associated with the Hoyle state but with higher-lying  states of that spin-parity. Indeed, consistent with {\it ab initio} observations, the NCSpM outcome demonstrates a large sensitivity of the energy of the Hoyle state and its $2^+$ excitation on the model space (Fig. \ref{gammadepend}a).
Finally, the additional degree of freedom associated with the $\gamma$ model parameter is in fact substantially limited by $0_2^+$ and $0_3^+$ and there is only a small window of reasonable $\gamma$ values (Fig. \ref{gammadepend}b),
where observables are also found in agreement with experiment: for  $\gamma$ from $10^{-2}$ to $10^{-5}$, e.g., $0^+_{g.st.}$ and $0^+_2$ rms radii increase 1.4 times, as well as $B(E2;\, 2^+_1 \rightarrow 0^+_{g.st.})$ and $Q_{2^+_1}$ increase four times.

In short, we carried forward a no-core shell-model study with a many-nucleon interaction to  further unveil the underlying physics behind various phenomena important to the low-energy nuclear dynamics of $^{12}$C. We showed, for the first time,  
how both collective states and states suggested to have cluster-like substructures 
emerge out of a fully microscopic, shell-model framework,
thereby providing a novel and essential perspective on the controversial Hoyle state.

\acknowledgments
We thank Pieter Maris, James P. Vary, David J. Rowe, Catherine M. Diebel, and Moshe Gai for useful discussions. 
This work was supported by the U.S. 
NSF (OCI-0904874), 
the U.S. DOE (DE-SC0005248 \& DE-FG02-95ER-40934), and the SURA.
ACD  acknowledges  support by the U.S. 
NSF (grant 1004822) through the REU Site in Dept. of Physics \& Astronomy at LSU. We acknowledge LONI
for providing HPC resources.

\end{document}